\documentclass[11pt]{article}
\def\beqar {\begin{eqnarray}}
\def\eeqar {\end{eqnarray}}
\def\beq {\begin{equation}}
\def\eeq {\end{equation}}
\def\ra {{\rangle}}
\def\la {{\langle}}
\def\half {{\textstyle{1\over 2}}}
\def\Tr {{\rm Tr}}
\def\tr {{\rm tr}}

\def\del {{\partial}}
\def\bdel{\bar{\partial}}

\def\bxi {{\bar \xi}}
\def\a {{\alpha}}
\def\b {{\beta}}
\def\g {{\gamma}}
\def\d {{\delta}}

\def\bi {{\bar i}}
\def\bj {{\bar j}}

\def \A {{\cal A}}
\def \D {{\cal D}}

\def \H {{\cal H}}

\def\no2 {{\textstyle{n\over 2}}}

\begin{document}

\begin{titlepage}
\null\vspace{-62pt}

\pagestyle{empty}
\begin{center}
\rightline{CCNY-HEP-04/2}
\rightline{March 2004}

\vspace{1.0truein} 
{\Large\bf  Edge states for quantum Hall droplets in higher dimensions }\\
\vskip .2in
{\Large\bf and a generalized WZW model}\\
\vskip .3in
\vspace{.5in}DIMITRA KARABALI $^{a}$
 and V.P. NAIR $^{b}$\footnote{E-mail addresses: karabali@lehman.cuny.edu (D. Karabali),
vpn@sci.ccny.cuny.edu (V.P. Nair)} \\
\vspace{.2in}{\it $^a$ Department of Physics and Astronomy\\
Lehman College of the CUNY\\
Bronx, NY 10468}\\
\vspace{.1in}{\it $^b$ Physics Department\\
City College of the CUNY\\  
New York, NY 10031}

\end{center}
\vspace{0.5in}

\centerline{\bf Abstract}
\vskip .1in
We consider quantum Hall droplets on complex projective spaces
with a combination of abelian and nonabelian
background magnetic fields.
Carrying out an analysis similar to what was done for abelian backgrounds,
we show that the
effective action for the edge excitations
is given by a chiral, gauged  Wess-Zumino-Witten (WZW)
theory generalized to higher dimensions.
\vskip .5in

\baselineskip=18pt

\end{titlepage}

\hoffset=0in
\newpage
\pagestyle{plain}
\setcounter{page}{2}
\newpage

\section{Introduction}

The study of quantum Hall effect in dimensions higher than
two and especially the edge excitations of quantum Hall droplets
have been
of  some research interest
following the original analysis by Zhang and Hu
\cite{HZ}. They considered the Landau problem for charged 
fermions on $S^4$ with a background magnetic field corresponding to 
the standard $SU(2)$ instanton.
In this case, the edge excitations, which correspond to the
deformations of the edge of a quantum Hall droplet, 
are gapless, as in two dimensions,
but now have spin
degrees of freedom as well.
In particular, one could have spin-two gapless excitations
suggesting an alternate description of the graviton.
Although
this has not been born out in the expected fashion, edge 
excitations for
a quantum Hall droplet lead to a very interesting class of
field theories.
The original model has by now been extended to many new 
geometries and contexts \cite{KN1}- \cite{NR}.
In a recent paper, we gave a general method for obtaining the
effective action for the edge excitations of a quantum Hall droplet
\cite{KN2}.
This was applied to higher dimensional quantum Hall systems,
specifically  for droplets in the complex projective space
${\bf CP}^k$. On ${\bf CP}^k$ one can have abelian or 
nonabelian background magnetic fields. In the case of an
abelian background, which was analyzed in
\cite{KN2}, the edge excitations can be described in terms of a chiral action for an abelian bosonic field. 
For ${\bf CP}^1=S^2$, this is the well known one-dimensional chiral bosonic action, defined on the boundary $S^1$, which describes the edge excitations of the $\nu = 1$ two-dimensional QHE \cite{SIKS}.  For general even dimensional spaces ${\bf CP}^k$, the notion of chirality is expressed by the fact that there is a preferred direction along the $S^{2k-1}$ droplet boundary which is relevant for the dynamics of the edge excitations. This is determined by the K\"ahler
form of ${\bf CP}^k$. The effective action is eventually integrated over all tangential directions. The bosonic field itself can be expanded in terms of the boundary surface oscillation modes of the incompressible droplet. This analysis also leads to the edge action
for QHE droplets on $S^4$, using the fact that ${\bf CP}^3$
is an $S^2$-bundle over $S^4$.

In this paper we extend our analysis of the edge effective action in the case of nonabelian background magnetic fields, thus completing
the full derivation of the edge action for all backgrounds.
\footnote {It may be worth noting that a particular case of the
action for edge states with a
nonabelian background was already considered in
the context of $S^3$ in \cite{NR}.} In this case the edge excitations are described in terms of a matrix valued bosonic field and the corresponding effective action is given in terms of a higher dimensional generalization of a nonabelian, chiral, gauged Wess-Zumino-Witten action. The form of the action suggests some kind of scale invariance and the possibility of an elegant current algebra structure. Therefore we believe these actions define an interesting new class of theories in higher dimensions which are worth studying in their own right.

This paper is organized as follows.
In section 2, we review briefly the results for the edge effective action for quantum
Hall droplets in ${\bf CP}^k$ with a $U(1)$ background field and state the main result 
that we 
derive for the nonabelian case.
In section 3 we present a detailed analysis when the background field is $U(k)$. 
The effective action for the edge excitations
is derived in the limit where the number of states of the lowest Landau level, $N$,
becomes very large. We start with  a discussion of the lowest Landau level
(LLL) wavefunctions. We then
discuss star products and evaluate the large $N$ limit of operator commutators in terms of symbols, which
are classical functions corresponding to operators. Using this we derive the large $N$ limit of the
kinetic energy term of the action. We separately evaluate the symbol corresponding to the density operator
and using this we simplify the potential energy term. We finally combine everything to arrive at the
effective action which is given in equation (\ref{4a}). 
This action is a higher dimensional generalization of a chiral, gauged  Wess-Zumino-Witten (WZW)
theory. In section 4 we discuss in more detail properties of this effective action and we point out
similarities to other higher dimensional WZW models, such as the K\"ahler WZW models \cite{nair-schiff}.
The paper is concluded with a short discussion.

\section{The effective action for edge states}

The effective action
for edge states was obtained in \cite{KN2} by considering fluctuations
of the density matrix following the strategy used by Sakita \cite{sakita}
in the case of the two-dimensional droplet.
We shall begin by briefly recalling the essential points of this method.
Let ${\hat \rho}_0$ be the density matrix corresponding to the droplet.
All elements of ${\hat \rho}_0$ are zero except for the diagonal ones corresponding to
the filled states, for which it is equal to one.
Time evolution of ${\hat \rho}_0$ is given by a unitary matrix ${\hat U}$. The
action which determines ${\hat U}$ is given by 
\beq
S= \int dt~ \left[ i \Tr ({\hat \rho}_0 { \hat U}^\dagger \del_t {\hat U})
~-~ \Tr ({\hat \rho}_0 {\hat U}^\dagger {\hat{\cal  H}} {\hat U}) \right]
\label{1}
\eeq
where $\hat {\H}$ is the Hamiltonian.
We used the following strategy in simplifying this action. First of all, we restrict
to the single particle Hilbert space, since it is a good approximation to neglect any
dynamically correlated states of two or more fermions. This is essentially a
Hartree-Fock approximation. For a droplet in the lowest Landau level (LLL) or any
fixed Landau level, we can then take the Hamiltonian,
up to an additive constant, as the potential
${\hat V}$ which confines the particles to the droplet.
If $N$ is the dimension of the one-particle Hilbert space
and $K$ states are occupied, we can simplify (\ref{1}) by taking $N\rightarrow \infty$
and $K$ very large but finite. 

We will first recapitulate the results for an abelian background. As we have shown in \cite{KN1},
\cite{KN2} the LLL states are characterized
by an integer $n\sim BR^2$ where
$B$ is the magnetic field and $R$ is the radius of the
${\bf CP}^k$; the dimension $N$ of the Hilbert space corresponding to the LLL sector depends on $n$, in
particular 
$N\sim n^k$ and scaling is done with $R\sim \sqrt{n}$. In the large $N$ limit two main results were used
to identify the edge effective action.

\begin{quotation}

1. The operator commutators tend to the Poisson brackets of appropriate classical functions, namely

\beq
[ \hat A , \hat B ] \rightarrow { i \over n} \{ A, B \}
\label{1a}
\eeq
where $A,~B$ are the classical functions corresponding to $\hat A,~\hat B$ and the Poisson bracket is
defined by
\beqar
\{ A, B\} &=& ({\Omega}^{-1})^{\bi j} \biggl( {\del A \over \del \bxi ^i}
{\del B \over \del\xi ^j} - {\del A \over \del \xi ^j}
{\del B \over \del\bxi ^i} \biggr) \nonumber\\
&=& i (1+\bxi \cdot \xi ) 
\Biggl( {\del A \over \del \xi^i}
{\del B\over \del \bxi^i} - {\del A \over \del \bxi^i }{\del B\over\del \xi^i}
 \nonumber\\
&&\hskip .2in+\xi\cdot {\del A\over \del\xi} 
\bxi \cdot {\del B\over\del\bxi} -\bxi\cdot {\del A \over
  \del\bxi}
\xi\cdot {\del B\over \del\xi}\Biggr)\label{2a}
\eeqar
\end{quotation}
Here $\xi ^i =(x_i + i y_i) / R,~ \bxi ^i = (x_i - i y_i)/ R$ are the local complex
coordinates and $\Omega$ is the K\"ahler two-form on ${\bf CP}^k$
\beqar
\Omega & = & \Omega_{i \bj} d\xi ^i \wedge d \bxi ^j \nonumber \\
 & = & i \biggl[ { {d\xi \wedge d\bxi} \over {1 + \bxi \cdot \xi}} - {{\bxi \cdot d\xi \wedge \xi \cdot
d\bxi} \over {(1 + \bxi \cdot \xi)^2}} \biggr]
\label{1c}
\eeqar

We can write ${\hat U}
= \exp (i {\hat \Phi})$ for some hermitian operator ${\hat \Phi}$,
expand various terms in (\ref{1}) in terms of commutators and extract their large
$N$-limit. If the background magnetic field is abelian, the classical field $\Phi$
corresponding to
${\hat
\Phi}$ is a scalar which parametrizes the surface deformations of the droplet. 
\begin{quotation}
2. The classical function $\rho_0$, the symbol corresponding to the density operator $\hat \rho_0 =
\sum_{i=1}^{K} \vert i \rangle \langle i \vert $ is essentially a step function identifying the droplet.
The term ${\del \rho_0 \over {\del r^2}}$, where $r^2 = \bxi \cdot \xi$, is a $\delta$-function with
support only at the edge of the droplet.
\end{quotation}

Using these two results and taking into account the scaling properties of various quantities as $n
\rightarrow \infty$, we found that the final effective action, up to an additive constant, is of the form
\beq
S = -{M^{k-1} \over {4 \pi^k}} \int_{\del\D} dt 
\left[ {\del \Phi \over \del t} + \omega ({\cal L}\Phi ) \right]
{\cal L}\Phi 
\label{3}
\eeq
where $M$ is related to the radius of the droplet, $R_D \sim \sqrt{M}$, and 
\beq
{\cal L}\Phi = i \biggl( \xi \cdot {\del \over \del \xi} - \bxi \cdot {\del \over \del \bxi} \biggr) \Phi
\label{4}
\eeq

Notice that the full effective action is an edge action defined on the boundary $\del \D$ of the droplet
$D$. ${\cal L}\Phi$ involves only derivatives of $\Phi$ with respect to a particular tangential 
direction, which is essentially determined by the K\"ahler form of the manifold. The density ${\rho}_0$
is chosen so as to minimize the potential energy, the particles being located at states of minimum
available energy.
$V$ is thus a constant along directions tangential
to the boundary of the droplet $\del \D$, it depends only on the 
normal coordinate $r$.
$\omega$ is a constant related to the derivative of the potential, in particular $\omega ={1 \over
n}{{\del V} \over {\del r^2}}$ (the scaling of $V$ is such that $\omega$ is $n$ independent).

Equation (\ref{3}) shows that the effective edge dynamics 
involves only the time-derivative of $\Phi$ and
one tangential derivative given by ${\cal L}\Phi$.
The action is chiral in
this sense.
Generally the field $\Phi$ depends on the 
remaining tangential
directions, leading to a
multiplicity of modes.
For a spherical droplet in even dimensions $2k$,
$\del\D \sim S^{2k-1}$. The mode analysis of the field $\Phi$ has been
indicated in \cite{KN2}.
In the special case of ${\bf CP^1}=S^2$, the corresponding action in (\ref{3}) is the well known
one-dimensional chiral bosonic action describing edge excitations for $\nu =1$ two-dimensional QHE
\cite{SIKS}.

On ${\bf CP}^k = SU(k+1)/U(k)$, it is
possible to have a nonzero background value for the
$SU(k)$ gauge fields
as well as the $U(1)$ fields.
As mentioned in the introduction, in this paper, we shall carry out a similar large $N$
simplification of the action for a background which has nonzero values for the
$SU(k)$ and $U(1)$ subalgebras. One of the main differences in the nonabelian case is that now the lowest 
Landau level states belong to a representation of $SU(k+1)$
with a lowest weight state
which transforms nontrivially under $SU(k)$, say, as a representation 
${\tilde J}$. As a result, the classical function $\Phi$ which is the symbol corresponding to the hermitian
operator
$\hat
\Phi$, is now a matrix of dimension $( dim \tilde{J} \times dim \tilde{J} )$. Again, as in the abelian
case, the key to extracting the large $N$ limit of the effective action are the two properties 1 and 2
mentioned above, appropriately modified in the case of the nonabelian bakground. 

Property 2 still remains true, with the difference that the symbol corresponding to the density operator
is a diagonal matrix of dimension $ (dim \tilde{J} \times dim \tilde{J})$ proportional to a step function,
which again identifies the droplet.

Property 1 is more seriously modified. The large $N$ limit of the
operator commutator, due to the matrix nature of the symbols,  will now involve a Poisson bracket-type
term along with a matrix commutator term. In particular
\beq
[ \hat A , \hat B ] \rightarrow [ A, B ] + { i \over n} \{ A, B \}_{gauged}
\label{4c}
\eeq
where
\beq
\{ A, B \}_{gauged} = ({\Omega}^{-1})^{\bi j} \bigl( D_{\bi}A D_j B - D_{\bi} B D_j A \bigr)
\label{4d}
\eeq
Notice that the Poisson bracket now involves the gauge covariant derivative $D = \del + [
\A,~~ ]$, where
$\A$ is the
$SU(k)$ gauge potential.

The relevant field in writing down the effective action is a unitary matrix
$G$ which is an element of
$U(dim {\tilde J})$.
As expected, since we have an $SU(k)$ background,
the action has an $SU(k)$  gauge symmetry. The field space
is essentially $U(dim{\tilde J})/SU(k)$.
The final result turns out to be a chiral, gauged Wess-Zumino-Witten action generalized to higher
dimensions.
\beqar
{\cal S}(G)&=& {1 \over {4 \pi^k}} M^{k-1} \int_{\del {\cal D}} 
dt ~ \tr \left[ \left( G^{\dagger} {\dot G} ~+~ \omega~G^{\dagger} {\cal L} G \right)
G^{\dagger}{\cal L}G \right] \nonumber\\ ~&&~+ (-1)^{{k(k-1)} \over 2}  {i\over 4\pi } {M^{k-1}\over
(k-1)!} 
\int_{\cal D} dt  ~2~ \tr \left[  
 G^{\dagger} {\dot G} (G^{-1}D G)^2 \right]\wedge \left({{i \Omega} \over \pi} \right)^{k-1}
\label{4a}
\eeqar
Here the first term is
on the boundary $\del {\cal D}$
of the droplet and it is precisely the gauged, nonabelian analogue of (\ref{3}). The operator ${\cal L}$ in
(\ref{4a}) is the gauged version of (\ref{4}),
\beq
{\cal{L}} = i \bigl( \xi^i D_i - \bxi^i D_{\bar i} \bigr)
\label{4b}
\eeq
The second term in (\ref{4a}), written as a differential form, is a higher dimensional
Wess-Zumino term; it is an  integral over
the rescaled droplet ${\cal D}$ itself, with the radial variable playing the role of the extra dimension.
This term is very similar to the Wess-Zumino term appearing in 
 K\"ahler-Wess-Zumino-Witten
models
\cite{nair-schiff}, used in the context of higher dimensional conformal field theories. The effective
action (\ref{4a}) is the main result of this paper.

\section{Analysis on ${\bf CP}^k$ with a nonabelian background field}
\vskip .1in
\noindent{1. $\underline{Wavefunctions}$}
\vskip .1in
We begin with a discussion of the Landau level states and wavefunctions
for the case when the background is nonabelian.
We will follow the group theoretic analysis given in \cite{KN1}, \cite{KN2}.
Since
${\bf CP}^k$ is the coset space ${SU(k+1) / U(k)}$,
the wavefunctions can be obtained
as functions on 
$SU(k+1)$ which have a specific transformation property under
the $U(k)$ subgroup.
Let $t_A$ denote the generators of $SU(k+1)$
as matrices in the fundamental representation; we normalize them
by $\Tr (t_A t_B )=\half \delta_{AB}$.
The generators corresponding to the $SU(k)$ part of
$U(k) \subset SU(k+1)$ will be denoted by
$t_a$, $a =1, ~2, \cdots , ~ k^2 -1$ and the
generator for $U(1)$
direction of the subgroup $U(k)$ will be denoted by 
$t_{k^2+2k}$. 
A basis of functions on $SU(k+1)$
is given by the Wigner $\cal{D}$-functions which are the
matrices corresponding to the
group elements in a representation
$J$
\beq
\D^{(J)}_{L;R}(g) = \la J ,L_i \vert {\hat g}\vert J, R_i \ra \label {5}
\eeq
where $L_i,~R_i$ stand for two sets of quantum numbers specifying the 
states on which the 
generators act, for left and right $SU(k+1)$ actions on $g$, respectively.
On an element $g\in SU(k+1)$, 
we can define left and right $SU(k+1)$ actions by
\beq
{\hat{L}}_A ~g = T_A ~g, \hskip 1in {\hat{R}}_A~ g = g~T_A
\label{6}
\eeq
where $T_A$ are the $SU(k+1)$ generators in the representation to which $g$ belongs.
Since ${\bf CP}^k = SU(k+1) /U(k)$, one can have, in general, a background 
magnetic field which is a $U(k)$ gauge field. Previously, we concentrated on the case
of a $U(1)$ field, here we are interested in the case of
a $U(k)$ background, where the wavefunctions transform as a specific
representation, say ${\tilde J}$, of $SU(k)$, for the action of the right generators 
$R_a \in SU(k)$ and carry
a charge corresponding to the right $U(1)$ generator $R_{k^2 +2k}$. 
The $U(1)$ generators $L_{k^2+2k}$ and $R_{k^2 +2k}$ will be referred to as
the ``hypercharge operators'', taking over the standard terminology for
$SU(3)$.

There are $2k$ right generators of $SU(k+1)$ which are not in
$U(k)$; these can be separated into $T_{+i}$, $i=1,2 \cdots ,k$, which are
of the raising type and $T_{-i}$ which are of the lowering
type. The covariant derivatives on ${\bf CP}^k$, in terms of 
their action on the wavefunctions,
can be identified with these ${\hat R}_{\pm i}$ right rotations on $g$. 
This is
consistent with the fact that the commutator of
covariant derivatives
is the magnetic field.
The commutators of ${\hat R}_{+i}$ and ${\hat R}_{-i}$ are in the
Lie algebra of $U(k)$; given the values of these $U(k)$ generators on
the wavefunctions, we see that they correspond to constant magnetic fields.
In the absence of a confining potential, the Hamiltonian ${\hat{\cal H}}_0$ may be 
reduced to the form 
$\sum_{i} {\hat R}_{+i} {\hat R}_{-i}$, apart from additive constants.
The corresponding energy eigenvalues of a charged particle on ${\bf CP}^k$ in the
presence of a nonabelian background are given by
\cite{KN1}
\beq
E = {1\over 2MR^2} \left[ C_2^{SU(k+1)}(J) - C_2^{SU(k)}({\tilde J})-
R^2_{k^2+2k}\right]
\label{7}
\eeq
where $M$ is the mass of the charged particles, $R$ is the radius of 
${\bf CP}^k$, $C_2$
is the quadratic Casimir operator for the group and representation indicated.

The states for a representation of $SU(k+1)$ can be labelled by two integers $(P,Q)$,
corresponding to a tensor of the form ${\cal T}^{\mu_1...\mu_Q}_{\nu_1...\nu_P}$ which is
symmetric in all the upper indices, symmetric in all the lower indices and traceless
for any contraction between upper and lower indices; the indices $\mu,~\nu$ take
values $1,2,\cdots,(k+1)$. In the case of a $U(1) \times SU(k)$ nonabelian background,
it is convenient to label the irreducible representation of $SU(k+1)_R$ by $(p+l,
q+l')$ corresponding to the tensor
\beq
{\cal T}^{\a_1...\a_q \g_{1}...\g_{l'}}_{\b_1...\b_p \d_{1}...\d_{l}} \equiv
{\cal T}^{q,l'}_{p,l} \label{7a}
\eeq
where $p,q$ indicate $U(1)$ indices and $l,l'$ indicate $SU(k)$ indices, namely
$\alpha$'s and $\beta$'s take the value $(k+1)$ and $\gamma$'s and $\delta$'s take
values $1,\cdots,k$. 

The right hypercharge corresponding to (\ref{7a}) is
\beq
\sqrt{2k(k+1)} R_{k^2+2k} = -k(p-q)+l-l' = -nk
\label{7b}
\eeq
The fact that $n$ has to be integer \cite{KN1} implies that $(l-l')/k$ is an integer,
thus constraining the possible $SU(k)_R$ representations $\tilde{J}$.

The single particle energy eigenvalues are now of the form
\beq
E = {1\over 2MR^2} \left[ C_2^{SU(k+1)}(p+l,q+l') - C_2^{SU(k)}(l,l')-
{{n^2 k} \over {2(k+1)}} \right] \label{7c}
\eeq
where
\beq
C_2^{SU(k+1)}(P,Q) = {k \over {2(k+1)}} \biggl[ P(P+k+1) +Q(Q+k+1) +{2 \over k} PQ
\biggr] \label{7d}
\eeq
Using (\ref{7b}), (\ref{7c}) and (\ref{7d}) we find that
\beq
E = {1\over 2MR^2} \left[ q^2 + q \bigl(l + l'+ {{l-l'} \over k}  +k
\bigr) +n \bigl(q+l+{k\over 2} \bigr) +l \bigl({{l-l'} \over k} + 1 \bigr) \right]
\label{7e}
\eeq
Since the background magnetic field and the representation (or 
charges) of the matter field are the given data for setting up the 
Landau problem, we must identify the wavefunctions for a given 
$SU(k)$ representation $\tilde{J}= (l,l')$ and $U(1)$ charge $-nk$. 
With $\tilde{J}$ and $n$ fixed, the lowest energy eigenstates 
correspond to $q=0$. The index $q$ plays the role of the Landau level index. Notice that for $\tilde{J}$ of
the form 
$\tilde{J}=(0, l')$, the lowest energy eigenstate is the lowest 
weight vector in the representation $J$ of $SU(k+1)$ annihilated by 
$\hat{R}_{i}$. This choice of background is particularly simple and for the 
rest of the discussion we will consider this case. (For other 
backgrounds, the effective action will have the same general 
qualitative features of the action we find here.) So the LLL states 
we consider correspond to the tensor ${\cal T}^{l'}_p$, where
$p=n-{l'
\over k}$ and $l'=jk,~j=1,2,\cdots$. \footnote{In deriving the LLL 
condition for ${\bf CP}^{2}$ in \cite{KN1} we had assumed $l > l'$ 
(the notation  $(k , k')$ was used instead of $(l, l')$). With this 
assumption the lowest energy states within the $q=0$ sector were of 
the form $(p+l, l)$. The assumption $l > l'$ is not necessary; 
without this assumption the lowest energy states are of the type 
$(p,l')$, with $l=0$.}

The Landau Hamiltonian ${\hat{\cal H}}_0$ commutes with the left action.
This corresponds to the magnetic translation symmetry of the Landau problem 
with the
left operators $\hat{L} _A$ representing magnetic translations. The degeneracy of the
LLL states corresponds to the dimension of the $J= (p,l')$ representation, where
\beq
dim^{SU(k+1)}(P,Q) = {{k (P+Q+k) (k+P-1)! (k+Q-1)!} \over {(k!)^2 P! Q!}}
\label{8a}
\eeq
Using $p=n-j$, $l'=jk$, $j=1,2,\cdots$, we find
\beq
dimJ = dim ^{SU(k+1)} (p,l') = {{k (n-j+jk+k) (k+n-j-1)! (k+jk-1)!} \over {(k!)^2
(n-j)! (jk)!}}
\label{8b}
\eeq
Similarly we find that
\beq
dim \tilde{J} = dim ^{SU(k)} (0,l') = {  {(k+jk-1)!} \over
{(k-1)!  (jk)!}}
\label{8c}
\eeq
Using (\ref{8b}), (\ref{8c}) we find that
\beq
dim J = dim \tilde{J} ~{{ (n-j+jk+k) (k+n-j-1)!} \over {k!
(n-j)!}}
\label{8d}
\eeq
In the thermodynamic limit $n \sim R^2 \rightarrow \infty$ we have 
\beq
c \equiv {{dim J} \over {dim \tilde{J}}} = {N \over N'} \rightarrow {n^k \over k!}
\label{8e}
\eeq
Following the analysis in \cite{KN1}, the completely filled LLL, $\nu =1$,
corresponds to a configuration of constant density
\beq
\rho _{\nu =1} \sim {N \over {N'~R^{2k}}} \rightarrow { n^k \over { k! R^{2k}}}
\rightarrow finite
\label{8f}
\eeq
In order to obtain a quantum Hall droplet for QHE on ${\bf CP}^k$ with a nonabelian
background, we need a $U(1)$ charge $n\sim R^2 \rightarrow \infty$ while the
dimension $N'=dim \tilde{J}$ of the $SU(k)$ representation remains finite.

The wavefunctions for the lowest Landau level are thus given by
\beqar
\Psi^J_{m; \alpha} (g) &=& \sqrt{N} ~\la J, L \vert ~{\hat g} ~\vert J, ({\tilde
J},\alpha ), - n\ra\nonumber\\ 
&=& \sqrt{N}~ {\cal D}^J_{m; \alpha}(g)\label{8}
\eeqar
where $N$ is the dimension of the representation $J$ of $SU(k+1)$.
The index $m=1,\cdots, N$ represents the state
within the $SU(k+1)$ representation $J$. The left generators $L_A$ act on these as linear transformations.
Further, the label $\alpha$ corresponds to the state $\vert \alpha\ra = \vert
J, ({\tilde J},
\alpha ), -n\ra$, $\alpha = 1,\cdots, N'$ and
\beqar
{\hat R}_a ~\Psi^J_{m; \alpha} (g) &=&
(T^{{\tilde J}}_a)_{\alpha \beta} \Psi^J_{m; \beta} (g) \nonumber\\ 
{\hat R}_{k^2 +2k} ~\Psi^J_{m; \alpha} (g) &=& - {n k\over \sqrt{2 k
(k+1)}}~\Psi^J_{m; \alpha} (g) \label{9}
\eeqar
The first of these equations shows that the wavefunctions (\ref{8}) under right
rotations transform as a  representation
of
$SU(k)$,
$(T^{{\tilde J}}_a)_{\alpha \beta}$ being the representation matrices for the
generators of
$SU(k)$ in the representation ${\tilde J}$.
$n$ is an integer characterizing the abelian part of the background field.
$\alpha ,\beta$ label states within the $SU(k)$ representation ${\tilde J}$
(which is itself
contained in the representation $J$ of $SU(k+1)$). The index $\alpha$ carried by the
wavefunctions (\ref{8}) is basically the gauge index. The wavefunctions are sections
of a
$U(k)$ bundle on ${\bf CP}^k$.
 
The wavefunctions (\ref{8})
are normalized by virtue of the orthogonality theorem
\beq
\int d\mu (g) ~\D^{*J}_{m;\alpha} (g)~\D^{J}_{m';\alpha '}
(g) ~=~ {\delta_{mm'}\delta_{\alpha \alpha '}\over N}
\label{10}
\eeq
where $d\mu (g)$ is the Haar measure on $SU(k+1)$, normalized to
unity. The orthonormality relations for wavefunctions should involve
integration over
the space, which is ${\bf CP}^k = SU(k+1) /U(k)$ rather than $SU(k+1)$ as we have
done in (\ref{10}). The difference has to do with the gauge transformations.
This will be clear if we consider $U(k)$-invariant combinations.
If $\phi^\alpha,~ \chi^\alpha$ 
transform
under $U(k)$ as the representation ${\tilde J}$, then the
combinations $\Psi^J_{m; \alpha} \phi^\alpha$ and $\Psi^J_{m; \alpha} \chi^\alpha$
are invariant and we can write
\beqar
\int d\mu ({\bf CP}^k) ~(\Psi^J_{m; \beta} \chi^\beta )^*
\Psi^J_{m; \alpha} \phi^\alpha &=& \int d\mu (g) ~
(\Psi^J_{m; \beta} \chi^\beta )^*
\Psi^J_{m; \alpha} \phi^\alpha \nonumber\\
&=& {\chi^\dagger\cdot \phi}\label{11}
\eeqar
The integration can be extended to the full group and the orthogonality theorem
(\ref{10}) used by virtue of the $U(k)$-invariance of the integrand.
\vskip .1in
\noindent{2. $\underline{Star ~products, ~commutators ~and ~Poisson ~brackets}$}
\vskip .1in
The simplification of the commutators in the large $N$-limit is achieved by use of the
star product which represents
the operator product in terms of classical functions 
corresponding to the operators. The classical function corresponding to an operator
is called the symbol of the operator.
To see how the symbol emerges naturally, consider the trace of a product of two
operators ${\hat A}$ and ${\hat B}$. By virtue of the orthogonality theorem, we
can write
\beq
c \sum_\alpha \int d\mu (g) {\cal D}^*_{m';\alpha} ~{\cal D}_{m; \alpha}
= \delta_{m'm}
\label{12}
\eeq
where $c = dim J / dim\tilde{J} = N/N' $.
Using this result we have
\beq
\Tr {\hat A} {\hat B} =  \sum_{ml} A_{ml} B_{lm} = c
\sum_{\alpha m l m'} \int d\mu (g) ~{\cal D}_{m;\alpha}~A_{ml} B_{lm'}~{\cal D}^*_{m';
\alpha}
\label{13}
\eeq
This shows that the appropriate definition of the symbol for $\hat{A}$ should be
\beq
A_{\alpha \beta}(g) = \sum_{ml}{\cal D}_{m;\alpha}(g) ~A_{ml} {\cal
D}^*_{l;\beta}(g)
\label{14}
\eeq
Notice that the symbol for an operator is now a $(N' \times N')$ matrix
valued function. Under a $U(k)$ transformation, $g \rightarrow g h$, it transforms  as
$A_{\alpha
\beta} 
\rightarrow h^T_{\alpha \gamma} A_{\gamma \delta} h^*_{\delta \beta}$ where $h$ is the
element of $U(k)$ in the matrix representation corresponding to $\tilde{J}$.
The formula for the Wigner function, namely equation (\ref{5}), shows that 
${\cal D}_{mn} (g^T) = {\cal D}_{nm}(g)$.  Thus the symbol for an operator ${\hat A}$ 
may also be written as
\beq
A_{\alpha \beta} (g) = \la \alpha \vert  {\hat g}^T {\hat A} {\hat g}^* \vert \beta \ra
\label{41}
\eeq

The symbol corresponding to the product of the operators 
${\hat A}$ and ${\hat B}$ is
\beqar
(\hat{A}\hat{B})_{\alpha \beta}(g) &=& \sum_{mrm'} {\cal D}_{m;\alpha}(g) ~A_{mr}B_{rm'}
{\cal D}^*_{m'; \beta}\nonumber\\
&=& \sum_{mrm's} {\cal D}_{m;\alpha}(g) ~A_{mr}~\delta_{rs}~B_{sm'}
{\cal D}^*_{m'; \beta}
\label{15}
\eeqar

The next step is to split the summation over the intermediate
indices using the completeness relation ${\bf 1} =\sum_A \vert A\ra \la A\vert$ where 
the summation is over all states in the $SU(k+1)$ representation. (The method outlined here for deriving
the star product has already been used in \cite{nair}, \cite{KN2}, \cite{NR}.)  The states can be grouped
into irreducible representations of $U(k)$. The first set of terms is $\sum_\alpha 
\vert J, ({\tilde J}, \alpha ) ,-n\ra
\la J, ({\tilde J}, \alpha ), -n\vert$. The next set of terms come from
$T_{+i} \vert J, ({\tilde J}, \alpha ), -n\ra = T_{+i} \vert \alpha \ra$. 
(We will abbreviate the state $\vert J, ({\tilde J}, \alpha ), -n\ra$ by
$\vert \alpha \ra$ when there is no likelihood of confusion.)
Since $T_{+i}$
transform as the fundamental representation of $SU(k)$, these states transform as the
product of
${\tilde J}$ and the fundamental representation. If desired one can split them into
irreducible components, but this is not necessary. The states
$T_{+i}\vert \alpha\ra$ form a complete set for the subspace
with $\sqrt{2 k (k+1)}~R_{k^2+2k} = -nk +k+1$. The scalar product is given by
\beqar
\la \alpha \vert T_{-i} T_{+j} \vert \beta\ra
&=& {n} ~\delta_{\alpha\beta} ~\delta_{ij}
- (T_a)_{\alpha\beta} ~f^a_{ij} \nonumber\\
&\equiv& {\cal G}(\alpha i, \beta j)
\label{15a}
\eeqar
where we have used the commutation rule
\beq
[ T_{-i} , T_{+j} ] = - \sqrt{ {2(k+1) \over k}}~ T_{k^2+2k}~ \delta_{ij}
- f^a_{ij}~ T_a\label{15b}
\eeq
$T_a$ is a generator of $SU(k)$. The completeness relation, writing out the
first two sets of states explicitly, becomes
\beq
{\bf 1} = \sum_\alpha \vert \alpha \ra \la \alpha \vert ~+~ 
\sum_{\alpha\beta ij} T_{+i} \vert \alpha \ra ~{\cal G}^{-1}(\alpha i, \beta j)
~\la \beta \vert T_{-j} ~+~\cdots \label{15c}
\eeq 
Putting in factors of $g$, this is equivalent to
\beq
\delta_{rs} = \sum_\gamma {\cal D}^*_{r,\gamma} {\cal D}_{s,\gamma}
- \sum_{\gamma \delta ij} {\cal G}^{-1}(\gamma i, \delta j) {\hat R}_{-j}{\cal
D}^*_{r,\delta} {\hat R}_{+i} {\cal D}_{s,\gamma} +\cdots
\label{15d}
\eeq
(The minus sign has to do with the fact that ${\hat R}_A g^\dagger = -T_A
g^\dagger$.) Since 
\beq
{\cal G}^{-1} (\gamma
i,
\delta j) = (1/n)
\delta_{\gamma\delta}
\delta_{ij} +{\cal O} (1/n^2) \label{15e}
\eeq
we can simplify (\ref{15d}) as
\beq
\delta_{rs} = \sum_\gamma \bigl( {\cal D}^*_{r,\gamma} {\cal D}_{s,\gamma}
-{1\over n} \sum_{i=1}^{k} {\hat R}_{-i}{\cal D}^*_{r,\gamma} {\hat R}_{+i} {\cal
D}_{s,\gamma}
\bigr) + {\cal O}\left( {1\over n^2}\right)
\label{16}
\eeq 
Using this in (\ref{15}) we find
\beqar
(\hat{A}\hat{B})_{\alpha \beta}(g) &=& \sum_{\gamma} \biggl( A_{\alpha \gamma}(g) B_{\gamma\beta}(g) 
-{1\over n}\sum_{i=1}^{k} {\hat R}_{-i}A_{\alpha\gamma}(g) {\hat R}_{+i} B_{\gamma\beta}(g)
\biggr) +{\cal O} \left( {1\over n^2}\right) \nonumber \\
& \equiv & \sum_{\gamma} A_{\alpha \gamma} * B_{\gamma \beta}
\label{17}
\eeqar
This is the definition of the star product.
The first term in the star product is thus the matrix product
$\sum_\gamma A_{\alpha \gamma} B_{\gamma \beta}$; this is in contrast to the case when
we have an abelian background for which we have functions as symbols and just a product
of functions as the leading term in the star product. 
Going back to (\ref{13}) we see that it can be written as
\beqar
\Tr {\hat A} {\hat B} &=& c \int d\mu (g)
\sum_{\alpha\beta} A_{\alpha\beta}* B_{\beta \alpha}\nonumber\\
&=& c \int d\mu (g) ~\tr (A*B)
\label{18}
\eeqar
where, in the second line, the trace (denoted by lower case tr)
is over the $SU(k)$
representation
${\tilde J}$ and $A, ~B$ are the symbols corresponding to ${\hat A},~ {\hat B}$.

For the symbol corresponding to the commutator of $\hat{A},~\hat{B}$, we have
\beq
([\hat{A}, \hat{B}]) (g) = [A(g), B(g)] -{1 \over n} \bigl( \hat{R}_{-i} A \hat{R}_{+i} B -
\hat{R}_{-i}B \hat{R}_{+i} A \bigr) + {\cal{O}} \left( {1 \over n^2}\right) \label{18a}
\eeq
The first term on the right hand side of (\ref{18a}) is a matrix commutator, given the fact that the
symbols $A(g),~B(g)$ are matrices. This term was absent in the case of the abelian background. Further in
the abelian case, the second term on the right hand side of (\ref{18a}) was found to be
proportional to the Poisson bracket $\{A,B\}$ \cite{KN2}. In the case of nonabelian background this
relation is more involved due to the presence of the nonabelian gauge fields, as we shall now discuss.

A convenient parametrization of the local complex coordinates of ${\bf CP}^k$ is given by
\beq
g_{i,k+1}= {\xi_i \over \sqrt{1+\bxi \cdot \xi}}, ~~i=1,\cdots,k~~\hskip .5in
g_{k+1,k+1} = {1\over \sqrt{1+\bxi \cdot \xi}}
\label{18b}
\eeq
where $g$ is an $SU(k+1)$ group element (fundamental representation). Since $g$ and $gh$, $h \in U(k)$,
are identified and correspond to the same point on ${\bf CP}^k = {{SU(k+1)}/ {U(k)}}$, we can use
the freedom of
$h$ transformations to write $g$ as a function of the coset coordinates $\xi^i, ~{\bar \xi}^i$ alone. We
can then write
\beq
g^{-1} dg = -it_{+i} E^{ i}_{ j} d \xi ^j -it_{-i} E^{ \bar i}_{ \bar j} d \bxi ^j-it_{a} E^{ a}_{
j} d \xi ^j-it_{a} E^{ a}_{\bar j} d \bxi ^j \label{18c}
\eeq
where $a$ takes values $1,\cdots,k^2$ and $k^2+2k$, and $i=1,\cdots,k$. The generators $t_A$ of the
$SU(k+1)$ algebra are divided into two sets, with $t_a$ belonging to the $U(k)$ algebra and the lowering
and raising coset operators
$t_{\pm i}$, such that $\tr (t_a t_{\pm i})=0$. (Notice that the components $E^i_{\bar j}$ and 
$E^{\bar i}_j$ are zero; this follows from (\ref{18b}) and is consistent with
the K\"ahler nature of ${\bf CP}^k$.)

The Cartan-Killing metric on ${\bf CP}^k$ is given by
\beq
ds^2 = E^i_j~ E^{\bar i} _{\bar l} ~d \xi^j~ d \bxi ^l
\eeq
which shows that the $E$'s are frame fields on the coset. The K\"ahler two-form on ${\bf CP}^k$ is likewise
written as
\beqar
\Omega & = & -i \sqrt{{2k \over {k+1}}} \tr \left( t_{k^2 + 2k} ~ g^{-1}dg \wedge g^{-1}dg \right)
\nonumber
\\ & = & i ~E^i_j~ E^{\bar i} _{\bar l} ~d \xi ^j \wedge d \bxi ^l ~\equiv ~\Omega _{j \bar{l}}~ d \xi ^j
\wedge d
\bxi ^l
\label{18d}
\eeqar

 From (\ref{18c}) we can write 
\beqar
(E^{-1})^j_i {{\del g} \over {\del \xi^j}} &=& g \left[-it_{+i} -it_{a} E^{ a}_{j} (E^{-1})^j_i \right]
\nonumber\\
(E^{-1})^{\bar j}_{\bar i} {{\del g} \over {\del {\bar\xi}^j}} &=& g \left[-it_{-i} -it_{a} E^{ a}_{\bar j}
(E^{-1})^{\bar j}_{\bar i}
\right]
\label{181}
\eeqar
We now define covariant derivatives on $g$ by
\beqar
D_i g  =  \del _i g - g {\cal A}_i ,~~~~~&&~~~~~D_{\bar i} g  =  \del_{\bar i} g - g {\cal
A}_{\bar i}\nonumber\\ 
{\cal A} _i  =  -i t_a E^a_i,~~~~~&&~~~~~{\cal A} _{\bar i}  =  -i t_a E^a_{\bar i}
\label{182}
\eeqar
Then (\ref{181}) becomes 
\beq
(E^{-1})^j_i D_j g = -i g ~t_{+i}, \hskip .3in (E^{-1})^{\bar j}_{\bar i} D_{\bar j} g = -i g ~t_{-i}
\label{183}
\eeq
Thus the right translation operators $\hat{R}_{\pm i}$ defined by $\hat{R}_{\pm i} g =g t_{\pm i}$ can be
identified as
\beq
\hat{R}_{+i} = i (E^{-1})^j_i D_j ~~~~~~~~~~~~~~~~~~~~\hat{R}_{-i} = i (E^{-1})^{\bar j}_{\bar i} D_{\bar
j} \label{184}
\eeq

The gauge field ${\cal A}=-it_a E^a$ is the potential corresponding to the background
magnetic field. It undergoes gauge
transformations under right
$U(k)$ rotations.
\beqar
g^{-1} dg & = & E + \A \nonumber \\
(gh)^{-1} d(gh) & = & h^{-1} E h + h^{-1} {\cal A} h + h^{-1} dh 
\label {185}
\eeqar
where $E$ and ${\cal A}$ are one-forms. Since $ h^{-1} dh~\in ~U(k)$, we find that the net effect is the
gauge transformation
\beq
{\cal A} \rightarrow h^{-1} {\cal A} h + h^{-1}dh ~~~~~~~~~~~~~~~~E \rightarrow h^{-1} E h \label{186}
\eeq

The symbol of an operator $\hat A$ has been defined as $A_{\a \b}(g) = \la \a \vert \hat{g} ^{T}
\hat{A} \hat{g} ^* \vert \b \ra $ in (\ref{41}); thus the action of the right operator $\hat{R}_{-i}$ on a
symbol is
\beqar
\hat{R}_{+i} A_{\a\b} & = & i (E^{-1})^j_i (D_j A)_{\a\b} \nonumber \\
D_j A & = & \del_j A + [{\cal A}_j,~ A]
\label{187}
\eeqar
where ${\cal A}_i = -i T_a E^a_i$ and $T_a$ are $U(k)$ generators in the $\tilde{J}$-representation.
Notice that the $U(1)$ part of the gauge field does not contribute in (\ref{187}). (A similar formula
holds for
${\hat R}_{-i}$.) Using (\ref{18d}) and (\ref{187}) we can now write
\beqar
\hat{R}_{-i} A \hat{R}_{+i} B - \hat{R}_{-i}B \hat{R}_{+i} A  & = & - (E^{-1})^\bj_\bi (E^{-1})^m_i
(D_\bj A D_m B - D_\bj B D_m A) \nonumber \\ &  = & - i( \Omega ^{-1}) ^{\bj m} (D_\bj A
D_m B - D_\bj B D_m A)
\label{188}
\eeqar
The final form for the symbol corresponding to the commutator of $\hat A,~\hat B$ in (\ref{18a}) is now
written as
\beq
([\hat{A}, \hat{B}]) (g) = [A(g), B(g)] +{i \over n} ( \Omega ^{-1}) ^{\bj m} (D_\bj A D_m B -
D_\bj B D_m A) + {\cal{O}} \left( {1 \over n^2}\right) \label{189}
\eeq
In terms of the complex coordinates $\xi,~\bxi$ we have
\beq
(\Omega ^{-1})^{\bj m} = -i (1 + \bxi \cdot \xi) ( \delta ^{jm} + \bxi ^j \xi ^m )
\label{190}
\eeq
In the case of an abelian background the covariant derivatives reduce to the regular derivatives and we
obtain the result in \cite{KN2}.

\vskip .1in
\noindent{3. $\underline{Simplification ~of ~the ~kinetic ~energy~term}$}
\vskip .1in
We will first consider the $N \rightarrow \infty$ limit of the kinetic term of the action
(\ref{1}). We start by writing ${\hat U} = \exp (i {\hat \Phi})$. The symbol
corresponding to
${\hat \Phi}$ is now a $(N' \times N')$-matrix $\Phi_{\alpha
\beta}$. In expanding
${\hat U}$ we encounter products of ${\hat \Phi}$'s, which have to be simplified by use
of star products. Consider for example ${\hat \Phi}^2$. The symbol corresponding to
this may be written as
\beq
\Phi * \Phi = \Phi~ \Phi - {1\over n} {\hat R}_{-i} \Phi~ {\hat R}_{+i} \Phi + \cdots
\label{19}
\eeq
where on the right hand side we have the matrix product of $\Phi$'s. (From now on we
assume that expressions with repeated indices involve a summation.) Let
$e_A$ form a basis for $(N' \times N')$ matrices, so that we can
write
$\Phi = e_A
\Phi^A$. The star product of the $\Phi$'s in (\ref{19}) can then be written as
\beqar
\Phi * \Phi &=& e_A e_B \left( \Phi^A - {1\over n} {\hat R}_{-i} \Phi^A
{\hat R}_{+i}\Phi^C {\del \over\del \Phi^C} \right) ~\left(\Phi^B - {1\over n}
{\hat R}_{-j}
\Phi^B {\hat R}_{+j}\Phi^D {\del \over\del \Phi^D} \right)\cdot 1 \nonumber\\
&=& {\tilde \Phi} ~{\tilde \Phi}\label{20}
\eeqar
where
\beqar
{\tilde \Phi} &=& e_A \left( \Phi^A - {1\over n} {\hat R}_{-i} \Phi^A {\hat
R}_{+i}\Phi^C {\del \over\del \Phi^C} \right)\nonumber\\
&=&e_A \left( \Phi^A + \xi^{AC} \del_C \right) 
\label{21}\\
\xi^{AC} &=&- {1\over n} {\hat R}_{-i} \Phi^A {\hat R}_{+i}\Phi^C\nonumber 
\eeqar
where $\del_C ={\del / \del \Phi^C} $.
It is easily checked that a similar formula holds for all higher powers of
${\hat \Phi}$'s, so that the simplification, at least to the leading $1/n$ order,
can be carried out in a closed form. (This way of simplifying operator products
is very close to the standard operator product expansions in field theory.
Although not needed for our purpose, the expansion can be carried out to any order
in $1/n$ by a similar formula.)
We now write the symbol corresponding to ${\hat U}$ as
\beqar
U&=& \exp (i {\tilde \Phi}) = \exp ( i e_A (\Phi^A + \xi^{AC}\del_C))\nonumber\\
&=& e^{i\Phi} ~+~ \int_0^1 d\alpha ~e^{i\Phi (1-\alpha)}
i e_A \xi^{AC} \del_C e^{i\Phi \alpha}~+\cdots\nonumber\\
&=& G ~-~ {i\over n} G~F + \cdots\label{22}
\eeqar
where
\beq
F = \int_0^1 d\alpha ~ e^{-i\alpha \Phi} ~{\hat R}_{-i} \Phi~
{\hat R}_{+i}(e^{i\alpha\Phi })
\label{23}
\eeq
and $G$ is the unitary $(N'\times N')$-matrix $e^{i\Phi}$.
In a similar way, we find
\beqar
U^\dagger &=& G^\dagger ~+~{i\over n} F^\dagger G^\dagger ~+\cdots\nonumber\\
F^\dagger &=& \int_0^1 d\alpha ~ {\hat R}_{-i} ( e^{-i\alpha \Phi})
~{\hat R}_{+i} \Phi~ e^{i\alpha \Phi} \label{24}
\eeqar
Using these results we find
\beq
U^\dagger * {\dot U} = G^\dagger {\dot G} -{1\over n} {\hat R}_{-i} G^\dagger ~
{\hat R}_{+i}{\dot G} - {i\over n} {\dot F} -{i\over n} G^\dagger {\dot G} F +{i\over
n} F^\dagger G^\dagger {\dot G} +\cdots
\label{25}
\eeq
where the overdot denotes differentiation with respect to time $t$.
The ${\dot F}$-term will give a total time-derivative upon taking the trace
(or integral) with $\rho_0$, so we shall drop it in what follows.
The term ${\hat R}_{-i} G^\dagger ~ {\hat R}_{+i}{\dot G}$ can be simplified as
\beqar
{\hat R}_{-i} G^\dagger ~ {\hat R}_{+i}{\dot
G} &=& {\hat R}_{+i} \left( {\hat R}_{-i}G^\dagger ~{\dot G}\right) - {\hat R}_{+i}
{\hat R}_{-i} G^\dagger ~{\dot G} \nonumber\\
&=& {\del \over \del t} \left( {\hat R}_{-i}G^\dagger ~{\hat R}_{+i}G\right) -
{\hat R}_{-i} {\dot G}^\dagger  {\hat R}_{+i} G \nonumber\\
&=& {\del \over \del t} \left( {\hat R}_{-i}G^\dagger ~{\hat R}_{+i}G\right) -
{\hat R}_{-i} \left( {\dot G}^\dagger  {\hat R}_{+i} G \right) +{\dot G}^\dagger
{\hat R}_{-i}{\hat R}_{+i} G\nonumber\\ 
&=& {1\over 2}{\del \over \del t} \left( {\hat R}_{-i}G^\dagger ~{\hat R}_{+i}G\right)
+{1\over 2} {\hat R}_{+i}( {\hat R}_{-i} G^\dagger {\dot G} ) - {1\over 2} {\hat
R}_{-i} ({\dot G}^\dagger {\hat R}_{+i} G ) \nonumber\\
&&\hskip .5in-{1\over 2} {\hat R}_{+i} {\hat R}_{-i} G^\dagger ~ {\dot G} 
+{1\over 2} {\dot G}^\dagger {\hat R}_{-i} {\hat R}_{+i} G
\label{26}
\eeqar
Derivatives on $G^\dagger$ can be written in terms of derivatives on $G$;
carrying this out for the relevant terms and rearranging,
the expression given above can be written as
\beqar
{\hat R}_{-i} G^\dagger ~ {\hat R}_{+i}{\dot G}&=&{1\over 2}{\del \over \del t} \left(
{\hat R}_{-i}G^\dagger ~{\hat R}_{+i}G\right) +{1\over 2} {\hat R}_{+i} ( {\hat R}_{-i}
G^\dagger {\dot G}) - {1\over 2} {\hat R}_{-i} ({\dot G}^\dagger {\hat R}_{+i}G )
\nonumber\\ && \hskip .5in +{1\over 4} \left[ G^\dagger {\dot G},~ ({\hat R}_{+i} {\hat
R}_{-i} G^\dagger ) G \right] -{1\over 4} \left[ G^\dagger {\dot G}, ~G^\dagger ({\hat
R}_{+i} {\hat R}_{-i} G)\right]
\nonumber\\
&&\hskip .5in -{1\over 4} \left[ G^\dagger {\dot G}, ~G^\dagger {\hat R}_{-i} G ~G^\dagger
{\hat R}_{+i} G + G^\dagger {\hat R}_{+i} G ~G^\dagger
{\hat R}_{-i} G \right]_+
\label{27}
\eeqar
where the square brackets with the subscript $+$ indicate the anticommutator
of the terms involved.

The expression for $F$ can be written as
\beqar
F&=& \int_0^1 d\alpha e^{-i\alpha \Phi } 
{\hat R}_{-i} \Phi ~{\hat R}_{+i} (e^{i\alpha \Phi})\nonumber\\
&=& \int_0^1 d\alpha~ {\hat R}_{+i} \left( e^{-i\alpha\Phi} {\hat R}_{-i} \Phi ~e^{i\alpha \Phi}
\right) - \int_0^1 d\alpha~ {\hat R}_{+i} \left( e^{-i\alpha \Phi} {\hat R}_{-i} \Phi \right)~
e^{i\alpha \Phi} \nonumber\\
&=& -i {\hat R}_{+i} (G^\dagger {\hat R}_{-i} G ) - \int_0^1 d\alpha~ {\hat R}_{+i} \left(
e^{-i\alpha \Phi} {\hat R}_{-i} \Phi \right)~ e^{i\alpha \Phi}\label{28}
\eeqar
In a similar way, $F^\dagger$ can be written as
\beq
F^\dagger = i {\hat R}_{+i} ( {\hat R}_{-i} G^\dagger )  ~G - \int_0^1 d\alpha~
{\hat R}_{+i} \left( e^{-i\alpha \Phi} {\hat R}_{-i} \Phi \right) e^{i\alpha \Phi}
\label{29}
\eeq
This shows also that
\beq
F- F^\dagger = -i G^\dagger {\hat R}_{-i} G ~ G^\dagger {\hat R}_{+i} G 
\label{30}
\eeq
We now rewrite the last set of terms in (\ref{25}) as
\beqar
i G^\dagger {\dot G} F - i F^\dagger G^\dagger {\dot G} &=&
{i\over 2} \left[ G^\dagger {\dot G}, ~F+F^\dagger \right]
+{i\over 2} \left[ G^\dagger {\dot G} ,~ F- F^\dagger \right]_+
\nonumber\\
&=&{i\over 2} \left[ G^\dagger {\dot G}, ~F+F^\dagger \right]
+{1\over 2} \left[ G^\dagger {\dot G},~ G^\dagger {\hat R}_{-i}G~ G^\dagger
{\hat R}_{+i}G \right]_+
\label{31}
\eeqar
Using (\ref{27}) and (\ref{31}), we can finally write
the kinetic term in the action as
\beq
K.E.= i c\int dt  d\mu ~\tr \rho_0 G^\dagger {\dot G} +i c \int dt d\mu \tr
\rho_0 \left[ G^\dagger {\dot G} , ~Q \right] ~+~c {\cal S}_1\label{32}
\eeq
where the antihermitian matrix $Q$ is given by 
\beq
Q = -{i\over 2n} (F+F^\dagger ) -{1\over 4n} ({\hat R}_{+i} {\hat R}_{-i} G^\dagger )
G +{1\over 4n} G^\dagger {\hat R}_{+i} {\hat R}_{-i} G\label{33}
\eeq
We have used the fact that $\sum_i [{\hat R}_{-i},~{\hat R}_{+i}] G 
=0$, since $\sum_i [{\hat R}_{-i},~{\hat R}_{+i}]$ is proportional to 
the (right) hypercharge.
The expression ${\cal S}_1$ is
given by
\beq
{\cal S}_1 = {i\over 2n}\int dt  d\mu ~\tr \rho_0 
\Biggl\{
{\hat R}_{-i} ({\hat R}_{+i} G^\dagger~ {\dot G} )  -
{\hat R}_{+i} ( {\dot G}^\dagger {\hat R}_{-i} G )  +{1\over 2}
\left[ G^\dagger {\dot G} ,~ [ G^\dagger {\hat R}_{+i} G , G^\dagger 
{\hat R}_{-i} G ] \right]_+ \Biggr\}
\label{34}
\eeq
(As mentioned before we have dropped the total time-derivative term.)

The first term in the kinetic energy term (\ref{32}) is a bulk term whose
interpretation we shall discuss shortly. The second term is also a bulk term.
It can be absorbed into the first term by a field redefinition.
Define ${\tilde G} = G \exp ( Q)$; we then find
\beq
i \int \tr \rho_0 {\tilde G}^\dagger {\dot {\tilde G}}
= i \int \tr \rho_0 G^\dagger {\dot G} +i \int \tr
\rho_0 \left[ G^\dagger {\dot G} , ~Q \right]
\label{35}
\eeq 
Thus the second term is absorbed into the first if we use ${\tilde G}$
in place of $G$. Since the change is of order $1/n$, 
we can do this in ${\cal S}_1$ as
well; it has the same form with $G$ replaced by ${\tilde G}$,
corrections being of order $1/n^2$.
The final answer for the kinetic term is then
\beq
K.E. = i c \int dt ~ d\mu ~ \tr \rho_0 G^\dagger {\dot G} ~+~ c {\cal S}_1 (G)
\label{36}
\eeq
where we have dropped the tildes on $G$, since we will be using this redefined 
$G$ from now on.

We now make some observations about the
scaling behaviour for various terms with $n$, which will be useful in the
ensuing discussion. The local coordinates
$\xi^i$ introduced in (\ref{18b}) are dimensionless
and we identify $R\xi^i$ as the coordinates
of the manifold, where $R$ is the radius of ${\bf CP}^k$. The ${\bf CP}^k$ measure of
integration $d\mu$ is written in terms of the coordinates $\xi$ as \cite{KN2}
\beqar
d\mu &=& {k! \over \pi ^k} {{d^k \xi d^k \bxi} \over {(1 + \bxi \cdot \xi)^{k+1}}}\nonumber\\
&=& i^k {k! \det \Omega \over \pi^k} ~d^k\xi ~d^k{\bxi}\label{44b}
\eeqar
Since $R\sim \sqrt{n}$, the
large $n$ limit is taken keeping
$\sqrt{n}~\xi$ finite. 
So a term like
$c \int d\mu $ will go like $c n^{-k}\sim 1$ as
$n\rightarrow \infty$ since $c\sim n^k$ as shown in (\ref{8e}). Such terms will be the
bulk terms in the action, for example, the term $\int \tr (\rho_0 
G^{\dagger}{\dot G})$.
For such terms, it is sufficient to retain the leading term of order one in
$\rho_0$, the subleading terms vanish as $n\rightarrow \infty$.
One can also have terms with two derivatives ${\hat R}_{+i}$ and ${\hat R}_{-i}$,
which arise with a power of $1/n$ in the action.
(These are typically boundary terms for the droplet as we shall see later.) 
The rescaling of these derivative terms will produce an
additional
power of $n$, so that a finite result is again obtained
with the leading term
in $\rho_0$. Thus for all terms in the action, it is sufficient to consider the
leading term in the density.   

The next step will be the simplification of the potential energy term. But before we
do this, we shall make a digression on how to evaluate the symbol 
$(\rho_0)_{\a\b}$ corresponding to the density
operator; this will help to simplify some of the calculations later.
\vskip .1in
\noindent{4. $\underline{A ~formula~for ~the~large~n-limit}$}
\vskip .1in

We now consider the symbol for the product ${\hat T}_B {\hat A} $ where
${\hat T}_B$ are the generators of $SU(k+1)$,
viewed as linear operators on the states
in the representation $J$. Using the formula (\ref{41}), it can be simplified as
follows.
\beqar
({\hat T}_B {\hat A} )_{\alpha\beta} &=& \la \alpha \vert~ {\hat g}^T~ 
{\hat T}_B ~{\hat A}~ {\hat g}^*
~\vert \beta \ra \nonumber\\
&=& S_{BC} ~\la \alpha \vert ~{\hat T}_C ~{\hat g}^T ~{\hat A}~ {\hat g}^* ~\vert \beta \ra
\nonumber\\
&=& S_{Ba} (T_a)_{\alpha\gamma}~ \la \gamma \vert ~{\hat g}^T ~{\hat A}~ 
{\hat g}^* ~\vert\beta\ra
+ S_{B+i} ~\la \alpha \vert~ {\hat T}_{-i}~ {\hat g}^T ~{\hat A}~ 
{\hat g}^* ~\vert \beta \ra\nonumber\\
&&\hskip .5in
+ S_{B~k^2+2k}~ \la \alpha \vert~ {\hat T}_{k^2+2k}~ {\hat g}^T~ {\hat 
A} ~{\hat g}^* ~\vert \beta \ra
\nonumber\\
&=& {\cal L}_{B\alpha \gamma}~ \la \gamma \vert ~{\hat g}^T ~{\hat A}~ 
{\hat g}^* ~\vert \beta\ra
\nonumber\\
&=& {\cal L}_{B\alpha \gamma}~ A(g)_{\gamma\beta}
\label{42}
\eeqar
where we have used ${\hat g}^T {\hat T}_B {\hat g}^* = S_{BC} {\hat T}_C$, $S_{BC} = 2 \Tr (g^T t_B g^* t_C)$.
(Here $t_B, t_C$ and the trace are in the fundamental representation of
$SU(k+1)$.)
${\cal L}_B $ is defined as
\beq
{\cal L}_{B\alpha\gamma}=
-\delta_{\alpha\gamma}{nk\over \sqrt{2k(k+1)}}S_{B ~k^2 +2k}~
~ +~\delta_{\alpha\gamma} S_{B+i} {\hat{{\tilde R}}}_{-i}
+S_{Ba} (T_a)_{\alpha\gamma}
\label{43}
\eeq
and ${\hat{{\tilde R}}}_{-i} $ is a differential operator
defined by $\hat{{\tilde R}}_{-i} g^T
= T_{-i} g^T $. (This can be related to ${\hat R}_{-i}$ but it is immaterial here.)
By choosing ${\hat A}$ as a product of $\hat{T}$'s, we can extend
the calculation of the symbol for any product of $\hat{T}$'s
using equation (\ref{42}). We find
\beq
({\hat T}_A {\hat T}_B \cdots {\hat T}_M)_{\alpha \beta}
= {\cal L}_{A\alpha\gamma_1} {\cal L}_{B\gamma_1 \gamma_2 }\cdots
{\cal L}_{M \gamma_r \beta} \cdot 1
\label{44}
\eeq
When $n$ becomes very large, the term that dominates in ${\cal L}_A$
is $S_{A~k^2+2k}$. These correspond to the coordinates 
of the ${\bf CP}^k$ space; $S_{A ~k^2+2k}$ obey algebraic constraints
which ensure that we are describing ${\bf CP}^k$ embedded in
${\bf R}^{k^2+2k}$. (The above relations also apply when the background is abelian
so that the lowest weight state is an $SU(k)$ singlet. In this case the
term with $(T_a)_{\alpha\beta}$ is absent in ${\cal L}_{A\alpha\beta}$.)
\vskip .1in
\noindent{5. $\underline{Evaluating~the~density}$}
\vskip .1in

The density is a function of the matrices $T_A$
in the representation $J$ of
$SU(k+1)$.
The method of evaluation of the symbol given above, eqs. (\ref{43})-(\ref{44}),
shows that its large $n$-limit corresponds to replacing
the $\hat{T}_A$'s by $-~(nk/\sqrt{2k(k+1)}~)S_{A~k^2+2k} ~\delta_{\alpha\beta}$. In
particular, the leading term of the density will be a
singlet under right
$SU(k)$. There will be nondiagonal subleading terms, which are down by a power of $n$, which
involve one power of the matrix $(T_a)_{\alpha\beta}$.

A more explicit analysis of the density will require more
details about the construction of the states. We found in section 3.1 that the LLL
states are essentially states of the $J$ representation of $SU(k+1)$ whose lowest
weight state is the $\tilde{J}$ representation of $SU(k)$. The lowest weight state
corresponds to the tensor ${\cal T}^{i_1 i_2\cdots i_{l'} 
}_{(k+1)(k+1)\cdots (k+1)} \equiv 
{\cal T}^{l'}_p $, where $i$'s take values $1, 2,\cdots ,k$ and all
the lower $p$ indices are of the $U(1)$ type and they are set to
$k+1$. This transforms as the symmetric rank $l'$ representation of $SU(k)$
and has hypercharge given by (\ref{9}) with $nk =pk +l'$.
This representation may also be written as the product ${\cal T}^{i_1 i_2
\cdots i_{l'}}\times {\cal T}_{(k+1)(k+1)...(k+1)}$. Going back to the notation for states
we used earlier, this can be written as
\beqar
\vert \alpha \ra \equiv \vert {i_1, i_2 ,...,i_{l'}};{(k+1),(k+1),...,(k+1)}\ra 
&=& \vert {i_1, i_2 ,...,i_{l'}}\ra\times \vert k+1\ra \times \vert k+1\ra \cdots
\times \vert k+1\ra\nonumber\\ &=& \vert {\tilde J}, \alpha \ra \times \vert k+1\ra
\times \vert k+1\ra \cdots \times \vert k+1\ra
\label{46}
\eeqar

%

At the next level, we have the states $T_{+i} ~{\cal T}^{i_1 i_2
...i_{l'}}_{(k+1)(k+1)...(k+1)}$ which are states of higher hypercharge in the
representation $J$.
We can obtain all the states of the representation $J$ by successive application of
$T_{+i}$'s. The states are naturally grouped into multiplets of $SU(k)$
with
increasing values of
hypercharge. 
 
We will now consider the simplification of the density 
when all the $SU(k)$ multiplets upto a fixed hypercharge are completely filled, starting from
the lowest, to form the droplet. This can be achieved by choosing
the potential to be linear on the hypercharge operator. In particular we choose it to be of the form
\beq
\hat{V} = \sqrt{2k \over k+1}~\omega ~\left( {\hat T}_{k^2+2k} + {nk \over
\sqrt{2k(k+1)}}
\right) \label{47}
\eeq
where $\omega$ is a constant. (The potential does not have to be exactly of this form;
any potential with the same qualitative features will do. We use this specific case
to carry out calculations explicitly.) The particular expression (\ref{47}) is such
that
\beq
\la s \vert \hat{V} \vert s \ra \label{47a}
\eeq
where $\vert s \ra$ denotes an $SU(k)$ multiplet of hypercharge $-nk +sk + s$, namely
$\sqrt{2k(k+1)}{\hat T}_{k^2+2k} \vert s \ra = (-nk +sk + s) \vert s \ra$.

When a number of $SU(k)$ multiplets up to a certain value of hypercharge is filled,
the density operator is
\beq
{\hat \rho_0}= \sum_{s=0}^M \vert s\ra \la s\vert \label{47b}
\eeq
The corresponding symbol for the density is
\beq
(\rho_0)_{\a\b} = \sum_{s=0}^M {\cal D}_{s;\alpha}(g) {\cal D}^*_{s;\beta}(g)
\label{47c}
\eeq
The first set of terms in $(\rho_0)_{\a\b}$ in (\ref{47c}), the term for $s=0$, is
given by
$\sum_\gamma
\la
\alpha
\vert \hat{g}^T
\vert \gamma\ra\la\gamma \vert \hat{g}^* \vert \beta\ra$. 
Using the fact that the lowest weight state is a product as in (\ref{46}),
we get
\beq
\sum_\gamma \la \alpha \vert {\hat g}^T
\vert \gamma\ra\la\gamma \vert {\hat g}^* \vert \beta\ra
= (g_{k+1,k+1}~g^*_{k+1,k+1})^p \la i_1...i_{l'}\vert \hat{g}^T\vert k_1...k_{l'}\ra
\la k_1...k_{l'}\vert \hat{g}^* \vert j_1...j_{l'}\ra
\label{48}
\eeq
where $g$ is a $(k+1) \times (k+1)$ matrix as in (\ref{18b}).
Further since the state $\vert j_1...j_{l'}\ra$ is itself a product
of states $\vert j_1\ra \vert j_2\ra...\vert j_{l'}\ra$ with suitable
symmetrizations, the term
$\la i_1...i_{l'}\vert \hat{g}^T\vert k_1...k_{l'}\ra
\la k_1...k_{l'}\vert \hat{g}^* \vert j_1...j_{l'}\ra$ is a product of terms like
$g^T_{i_1k_1}g^*_{k_1j_1}$. Adding and subtracting the term
$g^T_{i_1 (k+1)} g^*_{(k+1) j_1}$ and using unitarity of the $g$'s
we get
\beqar
g^T_{i_1k_1}~g^*_{k_1j_1} &=& \delta_{i_1j_1} - g^T_{i_1 (k+1)} g^*_{(k+1) j_1}
\nonumber\\
&=& \delta_{i_1j_1} ~+{\cal O}(1/n)
\label{49}
\eeqar
The second line in the above equation results from the fact that $g_{i(k+1)}\sim
\xi_i$, eq. (\ref{18b}), and the scaling  properties given after (\ref{36}). Further
since $l'$ is finite and $p \rightarrow n$ for large $n$, we find that (\ref{48}) can
be written as 
\beqar
(\rho)_{\alpha \beta}^{(s=0)}~=~\sum_\gamma \la \alpha \vert {\hat g}^T
\vert \gamma \ra \la \gamma \vert {\hat g}^* \vert \beta \ra
&=& (g_{k+1,k+1}~g^*_{k+1,k+1})^p ~~\delta_{\alpha \beta} \\
& \sim & {\d_{\a\b} \over {(1 + \bxi \cdot \xi)^n}}
\label{50}
\eeqar

Consider now the $s=1$ term in the expression for $(\rho_0)_{\a\b}$ in (\ref{47c}). It
is given by
\beq
\rho^{(s=1)}_{\alpha \beta}
=
\sum_{\gamma\delta i'j'} \la \alpha \vert \hat{g}^T~\hat{T}_{+i'} \vert \gamma
\ra ~{\cal{G}}^{-1}(\gamma i',
\delta j') ~\la \delta \vert \hat{T}_{-j'}~\hat{g}^* \vert \beta\ra
\label{51}
\eeq
When $T_{+i'}$ acts on $\vert \g \ra$, it can either act on the $U(1)$ indices of the
state converting them to $SU(k)$ indices or on the $SU(k)$ indices. The first
operation produces a factor $p$ since there are $p$ $U(1)$ indices, while the second
one produces a finite factor.  For large
$n$ the first set is dominant since $p \sim n$, so we can write 
\beqar
\la \alpha \vert {\hat g}^T~T_{+i'} \vert \gamma\ra
&=&
p (g_{(k+1) (k+1) })^{p-1}g_{(k+1)i'}
\la i_1...i_s\vert  \hat{g}^T \vert\gamma \ra  ~+ \cdots \nonumber\\
&\sim & n (g_{(k+1) (k+1) })^{n} \bxi_{i'}
\la i_1...i_s\vert  \hat{g}^T \vert\gamma \ra~+ \cdots 
\label{52}
\eeqar
This result then gives for large $n$
\beqar
\rho^{(s=1)}_{\alpha \beta}&=& n^2 (g_{k+1,k+1}~g^*_{k+1,k+1})^n
\bxi_{i'}~{\cal{G}}^{-1}(\gamma i',
\delta j') ~ \xi_{j'} \la \alpha\vert \hat{g}^T\vert \gamma\ra \la\gamma\vert \hat{g}^*
\vert\beta\ra~+ \cdots \nonumber\\
&=& n  \bxi\cdot\xi ~{\d_{\a\b} \over {(1 + \bxi \cdot \xi)^n}}
+ \cdots \label{53}
\eeqar
where (\ref{15e}) was used to simplify ${\cal{G}}^{-1}$ in (\ref{53}).
Notice that since $n \bxi\cdot \xi$ is fixed as $n\rightarrow \infty$, 
the first term
is
finite and the others are negligible at large $n$. Working out in a similar fashion the
large
$n$ contribution of the higher $s$ terms in $(\rho_0)_{\a\b}$ we find 
\beqar
(\rho_0)_{\alpha\beta}&=&
\left[ 1+ n \bxi\cdot\xi +\cdots
\right] {\d_{\a\b} \over {(1 + \bxi \cdot \xi)^n}} \nonumber\\
&=&\sum_{s=0}^{M} \left[{n! \over {s! (n-s)!}} \right] {{(\bxi
\cdot \xi)^s} \over {(1 + \bxi \cdot \xi)^n}} ~\delta_{\alpha\beta}
\label{54}
\eeqar
Thus the density is proportional to $\delta_{\alpha\beta}$ and the proportionality
factor is exactly the scalar density $\rho_0$ we obtained for the abelian background
\cite{KN2}. 
We thus get for the large $n$ limit 
\beqar
(\rho_0)_{\alpha\beta} &=&  \rho_0 ~\delta _{\alpha \beta}\nonumber \\
&\approx &
\Theta
\left( 1 -n{\xi\cdot \bxi \over M}\right)~\delta_{\alpha\beta}
\label{55}
\eeqar
The proof that the density for the abelian background is a step function for $N
\rightarrow \infty$ and large $M$ was given in detail in \cite{KN2}. The density in
(\ref{55}) corresponds to a nonabelian droplet configuration with boundary defined by
$n \bxi\cdot\xi = M$. The radius of the droplet is proportional to $\sqrt{M}$.
\vskip .1in
\noindent{$6.~\underline{Simplification ~of ~the ~potential~energy~term}$}
\vskip .1in
The large $n$ limit of the potential energy term can
now be derived in a fairly straightforward manner. First we calculate the symbol
corresponding to $\hat{V}$ as defined in (\ref{47}).
\beqar
V_{\alpha \beta} &=& \la \alpha \vert \hat{g}^{T} \hat{V} \hat{g}^{*} \vert \beta \ra
\\ &=& \omega \sqrt{{2k \over {k+1}}} \biggl[ \la \alpha \vert \hat{g}^{T} \hat{T}_{k^2
+ 2k} \hat{g}^{*} \vert \beta \ra + {{nk \d_{\a\b}} \over {\sqrt{2k (k+1)}}} \biggr] \label{55a}
\eeqar
Using (\ref{42}), (\ref{43}) we find
\beq
\la \alpha \vert \hat{g}^{T} \hat{T}_{k^2+2k}~ \hat{g}^{*} \vert \beta \ra = 
-\delta_{\alpha\beta}{nk\over \sqrt{2k(k+1)}}S_{k^2+2k, ~k^2 +2k}~
+S_{k^2+2k,a} (T_a)_{\alpha\beta}\label{55b}
\eeq
In the large $n$ limit the first term in the above equation dominates, so
combining (\ref{55a}) and (\ref{55b}) we find that the symbol for the potential is
\beq
V_{\alpha \beta} (g) = V(r) ~\delta_{\alpha \beta}  \label{55c}
\eeq
where $V(r)$ is the symbol for the potential in the case of the abelian
background \cite{KN2}. This depends only on the radial coordinate $r^2= 
\bxi \cdot
\xi$ and it was found in \cite{KN2} to be 
\beq
V(r) = \omega n {{\bxi \cdot \xi} \over {1 + \bxi \cdot \xi}} \label{55d}
\eeq

The potential energy term can be written as
\beqar
P.E. = \Tr {\hat\rho}_0 {\hat U}^\dagger {\hat V} {\hat U} &=& \sum_{l=0}^\infty
{(-i)^l \over l!} \Tr {\hat\rho}_0 [{\hat \Phi} , [{\hat \Phi} ,[ ...,[{\hat \Phi},
{\hat V}]...] \\
&=& \sum_{l=0}^\infty
{(-i)^l \over l!} \Tr [{\hat\rho}_0 , \hat{\Phi}] [{\hat \Phi} , [{\hat \Phi} ,[
...,[{\hat
\Phi}, {\hat V}]...]
\label{56}
\eeqar
where on the right hand side of (\ref{56}) there are $l~~\hat{\Phi}$'s involved. 
The first term in this sum, ($l=0$) is a constant field-independent term.
The term with $l=1$ is $\Tr [{\hat V}, {\hat \rho}_0 ] {\hat \Phi}$.
This is zero by the fact that $\rho_0$ is at the minimum of the potential.

Given the result (\ref{55}) for the density, the matrix commutator 
term $[\rho_0,~\Phi]=0$, so that
\beqar
([\hat{\rho}_0 , \hat{\Phi} ]) &=& -{1 \over n} \bigl( \hat{R}_{-i} \rho_0 \hat{R}_{+i} \Phi -
\hat{R}_{+i} \rho_0 \hat{R}_{-i} \Phi \bigr) \nonumber\\
&=& {i\over n} (\Omega^{-1})^{\bj m} \left( D_\bj \rho_0 D_m \Phi
- D_\bj \Phi D_m \rho_0\right)\label{57a}
\eeqar
where we used (\ref{189}).
The density $\rho_0$ is proportional to the identity, so its covariant derivative is the same as the
ordinary derivative; further it depends only on $r^2 =\bxi\cdot \xi$. We can then simplify the above result
as
\beq
([\hat{\rho}_0, \hat{\Phi}]) (g)
={1 \over n}~(1+\bxi \cdot \xi )^2 ~ {\del \rho_0 \over \del r^2} ~
\Biggl( \xi^i
~D_i \Phi - \bxi^i ~D_\bi \Phi \Biggr)\label{57b}
\eeq
Taking into account the scaling of the coordinates $\xi$ we find that the large $n$
limit of the above expression is
\beq
([\hat{\rho}_0, \hat{\Phi}]) (g) = -{i \over n} {{\del \rho_0} \over {\del r^2}} {\cal{L}}
\Phi\label{57c}
\eeq
where
\beq
{\cal{L}} = i \bigl( \xi^i D_i - \bxi^i D_{\bar i}
 \bigr) \label{58}
\eeq
Similarly
\beq
([\hat{\Phi}, \hat{V}]) (g) = {i \over n} {{\del V} \over {\del r^2}} ~{\cal{L}} \Phi\label{58a}
\eeq

The leading order term in the potential term (\ref{56}) as $n$ becomes large can now be
written as
\beq
P.E. =\sum_{l=2}^{\infty}
{(-i)^l \over l!}{c \over n^2}\int \tr \left[ {\del \rho_0 \over \del r^2} ~{\cal
L}\Phi ~\underbrace { [\Phi ,[\Phi ,...[\Phi }_{l-2} ,{\cal L}\Phi ] ...]]~
{\del V
\over
\del r^2}
\right] +\cdots
\label{59}
\eeq 

Consider now the expression $\tr ( G^{\dagger} {\cal L} G )^2$.
We can write this as
\beqar
\tr ( G^{\dagger} {\cal L} G )^2&=& \tr \left[ i \int_0^1 d\alpha
e^{-i\alpha \Phi} {\cal L}\Phi ~e^{i\alpha \Phi} ~i\int_0^1
d\beta e^{-i\beta \Phi } {\cal L}\Phi~ e^{i\beta \Phi} \right]\nonumber\\
&=& -\tr \left[  \int_0^1 d\alpha d\beta
 {\cal L}\Phi ~e^{i(\alpha -\beta ) \Phi} 
{\cal L}\Phi~ e^{-i(\alpha -\beta ) \Phi} \right]\nonumber\\
&=& \tr \left[ \int_0^1 d\alpha \int_\alpha^{\alpha-1} d\gamma
\sum_{l=0}^\infty {(i\gamma )^l\over l!} {\cal L}\Phi 
\underbrace{[ \Phi , ...,[\Phi}_l ,{\cal L}\Phi ]...] ~\right]\nonumber\\
&=& 2 \sum_{l=2}^\infty \tr \left[{i^l \over l!} {\cal L}\Phi \underbrace{[\Phi ,...,
[\Phi}_{l-2} , {\cal L}\Phi ]...] ~\right]
\label{60}
\eeqar
There are no odd $l$ terms in (\ref{60});
the odd $l$ terms in (\ref{59}) are also zero by
trace identities since the density $(\rho_0)_{\a\b}$ and the potential $V_{\a\b}$ are
proportional to the identity matrix. Combining equations (\ref{59}), and (\ref{60}) we
get
\beqar
P.E.
&=& \sum_{l=2}^\infty
{(-i)^l \over l!}{c\over n^2}\int \tr \left[
{\del \rho_0 \over \del r^2} ~{\cal L}\Phi ~ [\Phi ,[\Phi ,...[\Phi ,{\cal L}\Phi ]
...]]~
{\del V
\over
\del r^2}
\right] +\cdots \nonumber\\
&=& {c\over 2n^2} \int {\del \rho_0 \over \del r^2}
~\tr (G^{\dagger}{\cal L} G )^2 ~{\del V\over \del r^2}~+\cdots\nonumber\\
&=& {{\omega c} \over 2n}\int {\del \rho_0 \over \del r^2}~
\tr (G^{\dagger}{\cal L} G )^2~+~\cdots
\label{61}
\eeqar

This result is for densities which behave as $\delta_{\alpha\beta}$ at large $n$. In
the last line of (\ref{61}) $\rho_0$ is the scalar density for the abelian background
case as discussed after (\ref{54}).
\vskip .1in
\noindent{7. $\underline{The ~effective ~action}$}
\vskip .1in
We are now in a position to combine the results for the kinetic and potential energy
terms. The kinetic term (\ref{36}) simplifies further if we use the fact that
$(\rho_0)_{\a\b} \sim \d_{\a\b}$.  The kinetic energy term is $K.E. = c {\cal S}_1(G)$, where 
${\cal S}_1(G)$ is given
in  (\ref{34}) and can be written as
\beqar
K.E. &=& {ic\over 2n} \int \tr  G^\dagger {\dot G}\left(
{\hat R}_{-i}\rho_0 ~G^\dagger {\hat R}_{+i} G - 
 G^\dagger {\hat R}_{-i}G ~{\hat R}_{+i} \rho_0 \right) +{{ic}\over 2n} \int \rho_0 \tr
\left(  G^{\dagger} {\dot G}  \bigl[ G^{\dagger} {\hat R}_{+i} G , 
~G^{\dagger} 
{\hat R}_{-i} G \bigr]  \right)\nonumber\\
&=& -{c\over 2n} \int {\del \rho_0 \over \del r^2} ~ \tr
(G^\dagger {\dot G} ~G^\dagger {\cal L} G ) +{{ic}\over 2n} \int \rho_0  \tr
\left(  G^{\dagger} {\dot G}~ \bigl[ G^{\dagger} {\hat R}_{+i} G, ~ 
G^{\dagger} 
{\hat R}_{-i} G \bigr]  \right)
\label{62}
\eeqar
In deriving the second line in (\ref{62}) we used (\ref{57a}), (\ref{57b}).
We can write this part of the action in a different form by using (\ref{188}).
\beqar
K.E. &=& {c\over 2n}\int dt d\mu~ (\Omega^{-1})^{\bj m}
\tr \Biggl[ G^{-1}{\dot G}\biggl(  D_\bj \rho_0 ~G^{-1} D_m G
-~D_m \rho_0 ~G^{-1} D_\bj G \nonumber\\
&&\hskip 1.2in 
- \rho_0 \biggl( ~G^{-1} D_\bj G~ G^{-1} D_m G - G^{-1} D_m G~ G^{-1} D_\bj G \biggr) \biggr) \Biggr]
\label{62a}
\eeqar
The volume element of ${\bf CP}^k$ given in (\ref{44b})
has a factor $\det \Omega$; using the relation
\beq
\epsilon^{i_1 i_2 ...i_{k-1} m}
\epsilon^{{\bar j}_1 {\bar j}_2 ...{\bar j}_{k-1} \bj}
\Omega_{i_1 {\bar j}_1}\Omega_{i_2 {\bar j}_2}...\Omega_{i_{k-1} {\bar j}_{k-1}}
= (k-1)!~(\det \Omega ) (\Omega^{-1})^{\bj m} 
\label{62b}
\eeq
we can now write the kinetic term as
\beqar
K.E. &=& {{i^k k c} \over {2n \pi^k}} \int  dt~\tr
\Biggl[ G^{-1}{\dot G} \biggl(  \del_\bj \rho_0 ~G^{-1} D_m G
-~\del_m \rho_0 ~G^{-1} D_\bj G 
- ~\rho_0 G^{-1} D_\bj G~ G^{-1} D_m G \nonumber\\
&&\hskip 0.5in + \rho_0 G^{-1} D_m G~ G^{-1} D_\bj G \biggr) \Biggr]
\times\epsilon^{i_1 ...i_{k-1} m}
\epsilon^{{\bar j}_1 ...{\bar j}_{k-1} \bj}
\Omega_{i_1 {\bar j}_1}...\Omega_{i_{k-1} {\bar j}_{k-1}} d^k \xi d^k \bxi\nonumber\\
&=& {{i k c} \over {2n \pi}} (-1)^{{k(k-1)} \over 2}\int dt ~\tr
\Biggl[ G^{-1}{\dot G} \biggl(- d\rho_0 G^{-1}DG  + \rho_0 G^{-1}DG ~G^{-1}DG\biggr)\Biggr]\wedge
\left( {i \Omega\over \pi}\right)^{k-1}
\label{62c}
\eeqar
where $\Omega$ is the symplectic two-form defined in (\ref{1c}).
The last line in (\ref{62c}) has been written in terms of differential forms. 

Combining this with
with the potential term (\ref{61}), 
our final result for the effective action becomes
\beqar
{\cal S}(G)
&=& -{c \over 2n} \int  dt d\mu ~{{\del \rho_0} \over \del r^2} ~
\tr \left[ \left( G^{\dagger} {\dot G} + \omega ~G^{\dagger} {\cal L} G \right)  
G^{\dagger}{\cal L}G  \right] \label{63}\\
&&\hskip .3in
+{ikc\over 2\pi n} (-1)^{{k(k-1)} \over 2} \int_{\cal D} dt ~\rho_0 \tr \left[ G^{-1}{\dot G} 
 G^{-1}DG ~G^{-1}DG \right] \wedge
\left({{i \Omega} \over \pi} \right)^{k-1}\nonumber
\eeqar
We have used both forms (\ref{62}) and ({\ref{62c}).

${\cal S}(G)$ is a generalized chiral, gauged Wess-Zumino-Witten (WZW) action \cite{WZW}, \cite{gWZW}.
Since $\rho_0$ is a step function as in (\ref{55}), its derivative with respect to 
$r^2$ is a delta function with support at the boundary $\del {\cal D}$
of the droplet, namely
\beq
{{\del \rho_0} \over {\del r^2}} = -{n\over M} ~\d \left( 1 - {{n r^2} \over M}
\right)
\label{63a}
\eeq
Thus the first term on the right hand side of
(\ref{63}) is evidently a boundary term.
The second term
is a (gauged) Wess-Zumino term for the field
$G$; the integration is over the droplet whose boundary is the edge $\del {\cal D}$. 
Usually the Wess-Zumino term is written in terms of integration
over a three-manifold whose boundary is the spacetime of interest;
in such a representation, the Wess-Zumino term
can be displayed as the integral of a local differential form.
Here we see that the droplet,
along with time,
plays the role of this three-manifold, the radial variable of the droplet
serving as the extra
dimension. Of
course this is not really a three-manifold, the actual dimension is $2k+1$.
Only derivatives with respect to a particular spatial direction,
namely ${\cal L}$, appear in the action. However, once the integration
over the various directions are carried out the result will involve all
the coordinates of the edge of the droplet for the first term.
Finally, since the gauged Wess-Zumino term is the integral of a locally 
exact form, as shown in the next section,
it should be considered as part of the boundary action;
it is not a bulk term.

\section{Properties of the effective action}
\vskip .1in
\noindent{$1.~\underline{Gauge ~symmetry}$}
\vskip .1in
The effective action we have derived is a generalized gauged WZW action.
The field $G$ is a unitary $(dim {\tilde J}\times dim{\tilde J})$-matrix
and it is gauged vectorially with respect to the group $SU(k)$.
As we have mentioned earlier, the $U(1)$ gauge field does not appear in the covariant derivative and in
the effective action.

Usually, in a gauged WZW action, the kinetic terms have gauge 
covariant derivatives 
replacing the ordinary
derivatives as in (\ref{63}). However, the gauging of the WZ-term is 
not done by replacing
derivatives by covariant derivatives, rather it involves 
additional terms which are local boundary terms such that the anomalies
of the gauge symmetry cancel out \cite{gWZW}.
Explicitly, such a gauged WZ-term with a vector gauge symmetry
is given, for a three-dimensional manifold, by
\beq
\Gamma_{WZ}(G, {\cal A})= -{1\over 12\pi} \int \tr \Biggl[ 
(G^{-1}dG)^3 + 3 d \left( {\cal A}~ dG G^{-1} +{\cal A}~  G^{-1} dG + {\cal A} G {\cal A} G^{-1}\right)
\Biggr]
\label{prop1}
\eeq
The WZ-term in the effective action (\ref{63}) has gauge covariant 
derivatives; it is of the form $\tr (G^{-1}DG)^3\wedge (\Omega)^{k-1}$,
with ${\cal A}_0 =0$. (${\cal A}_0 $ is the potential corresponding to the time-direction.)
Expanding out the covariant derivatives we find
\beqar
-{1\over 12\pi}\tr (G^{-1} DG)^3 &=& \Gamma_{WZ}(G, {\cal A}) -{1\over 4\pi}
\tr\left[ (dG G^{-1} +G^{-1}dG ) {\cal F}\right]\nonumber\\
&=& \Gamma_{WZ}(G, {\cal A}) -{1\over 4\pi}
\tr\left[ ({\dot G} G^{-1} +G^{-1}{\dot G} ) {\cal F}\right]\nonumber\\
&=& \Gamma_{WZ}(G, {\cal A}) -{1\over 4\pi}
\tr\left[ ({\dot G} G^{-1} +G^{-1}{\dot G} ) T_a\right] (-i{\cal F}^a)
\label{prop2}
\eeqar
Here ${\cal F}$ is the field strength associated to the potential ${\cal A}$
and in the second line we use the fact
that ${\cal F}$ has no time-components.
The field strength ${\cal F}$ can be calculated from the potential as
\beq
{\cal F}^a = 2i \tr ( t^a [t_{+i} , t_{-j}]) ~ E^i_k E^{\bar j}_{\bar l}
d\xi^k \wedge d\bxi^{\bar l}
\label{prop3}
\eeq
The wedge product with $d\mu ({\bf CP}^{k-1})$ in (\ref{prop2})
 leads to the term
${\cal F}^a_{{\bar l}r} (\Omega^{-1})^{{\bar l}r}$.
This is verified to be zero using 
$(\Omega^{-1})^{{\bar l}r}= -i (E^{-1})^{\bar l}_{\bar m} (E^{-1})^r_m$
and the fact that $\sum_i \tr (t_a [t_{-i} , t_{+i}])=0$.
Thus $(G^{-1}DG)^3$ does indeed reduce to the usual gauging of the
WZ-term for the present case.

{\it In summary, we have a gauged WZW model generalized to higher dimensions
where the field $G$ takes values in the unitary group
$U(dim {\tilde J})$ and the gauge group is $SU(k)$, in other words
a $U(dim {\tilde J}) /SU(k)$ coset model. }
\vskip .1in
\noindent{$2.~\underline{The ~level ~number ~of ~the ~action}$}
\vskip .1in
The WZ-term in (\ref{63}) has the time-derivative term separated off.
We can write it in a more symmetric fashion as
\beqar
{\cal S}_{WZ} &=& {ikc\over 2\pi n} (-1)^{{k(k-1)} \over 2} \int_{\cal D} dt \rho_0 \tr \left[ G^{-1}{\dot
G} 
 G^{-1}DG ~G^{-1}DG \right] \wedge
\left( {i \Omega \over \pi} \right) ^{k-1}\nonumber\\
&=& {ikc\over 6\pi n} (-1)^{{k(k-1)} \over 2} \int_{\cal D} \rho_0 \tr  
 (G^{-1}DG)^3 \wedge
 \left( {i \Omega \over \pi} \right) ^{k-1} \nonumber\\
&=& {ikc\over 12\pi n} (-1)^{{k(k-1)} \over 2} \int_{\cal D} \rho_0 ~2~\tr  
 (G^{-1}DG)^3 \wedge
\left( {i \Omega \over \pi} \right) ^{k-1}
\label{prop4}
\eeqar
In comparing this with the usual WZ-term, we have to take account of an additional factor
of $\half$ which arises due to the fact that $r^2$, not $r$, is the variable of integration
for the `extra coordinate'. (The volume element will have a factor
$rdr$; with the factor of $2$ we inserted inside the
integral, this will become $d(r^2)$.) Equation (\ref{prop4})
shows that the level number of the WZ-term is
given by
\beqar
{c k \over n} &=& {dim J \over dim {\tilde J}} ~{k\over n}\nonumber\\
&=& \left[ {(p+k-1)! \over p! (k-1)!}\right] ~{p+k +jk \over {p+j}}\nonumber\\
&\rightarrow& \left[ {(p+k-1)! \over p! (k-1)!}\right]
\label{prop5}
\eeqar
as $p\rightarrow \infty$. The level number is indeed an integer, as it should be,
 and is very large
as $p\rightarrow \infty$. 

Now that we have established this property, we can rewrite the action in another
more transparent form.
Using (\ref{55}), (\ref{63a}), (\ref{8e}), (\ref{44b}) and the $n$-scaling of the
coordinates we can further simplify (\ref{63}) as 
\beqar
{\cal S}(G)&=& {1 \over {4 \pi^k}} M^{k-1} \int_{\del {{\cal D}}} 
\tr \left[ \left(G^{\dagger} {\dot G} + \omega ~G^{\dagger} {\cal L} G \right) 
G^{\dagger}{\cal L}G \right] \nonumber\\
~&&~+ (-1)^{{k(k-1)} \over 2} {i\over 4\pi } {M^{k-1}\over (k-1)!}  \int_{\cal D} 
~2~ \tr \left[  
 G^{\dagger} {\dot G} (G^{-1}D G)^2 \right]\wedge \left({{i \Omega} \over \pi} \right)^{k-1}
\label{63b}
\eeqar
The right hand side of this equation is entirely in terms of rescaled coordinates,
with the radius of the droplet being unity.
We have taken the limit of large $n$ for the prefactors as well.
Thus the
terms displayed here are the terms which have a finite limit as $n \rightarrow \infty$;
the corrections to this action are of order $1/n$.
\vskip .1in
\noindent{$3.~\underline{Chirality}$}
\vskip .1in
The effective action is a gauged WZW model and carries a notion of chirality \cite{SS}.
It has the structure
$~\tr \left[ \left( G^{\dagger} {\dot G} ~ ~+~
\omega ~G^{\dagger}{\cal L}G \right)~G^{\dagger} {\cal L} G\right]$; the 
combination $G^{\dagger} {\dot G} ~ ~+~
\omega ~G^{\dagger}{\cal L}G$ indicates that 
there is 
chirality in the theory. The eventual integration
over all spatial directions will give an action which has 
rotational symmetry, but the time-derivative
will still occur in a similar combination.
\vskip .1in
\noindent{$4.~\underline{Relationship ~to ~other ~higher ~dimensional ~WZW ~models}$}
\vskip .1in
The effective action we have obtained is a higher dimensional generalization of the
WZW action. Going back to (\ref{62c}), we write it as
\beq
{\cal S}(G) ={ikc\over 2\pi n} (-1)^{{k(k-1)} \over 2}\int \tr
\Biggl[ G^{-1}{\dot G} \biggl(- d\rho_0 G^{-1}DG  + \rho_0 G^{-1}DG ~G^{-1}DG\biggr)\Biggr]\wedge
\left( {i\Omega\over \pi}\right)^{k-1}
\label{prop6}
\eeq
The action density which is integrated is in the form of the exterior product of
the density for the
two-dimensional WZW action with $\Omega^{k-1}$. A generalization of the 
WZW action to higher dimensions along these lines has been considered before
\cite{nair-schiff}. The action for a $k$-complex dimensional
K\"ahler manifold is
\beq
{\cal S}(G) = -{1\over 2\pi} \int \tr ( G^{-1} \del G ~G^{-1}\bdel G)\wedge \Omega^{k-1}
+{i\over 12\pi} \int \tr (G^{-1}dG)^3 \wedge \Omega^{k-1}
\label{prop7}
\eeq
For $k=2$, this was obtained as the boundary action for a K\"ahler-Chern-Simons theory
for a gauge field $A$ with an action of the form
\beq
{\cal S} = -{k\over 4\pi} \int \tr \left( A dA +{2\over 3} A^3 \right)\wedge \Omega
~+~\int \tr (\Phi F +{\bar \Phi} F)
\label{prop8}
\eeq
This action is defined on ${\cal M}\times [t_i ,t_f]$
where
${\cal M}$ is a K\"ahler manifold and $[t_i,t_f]$ is an interval of time.
$\Phi$ is a Lie algebra-valued $(2,0)$ form on ${\cal M}$
and a one-form for the time direction, i.e., it
is of the form ${\half}\phi_{ij} d\xi^i d\xi^j dt$.
This theory has the equations of motion
\beqar
F\wedge \Omega &=& 0\nonumber\\
F_{ij} = F_{{\bar i} {\bar j}} &=&0
\label{prop9}
\eeqar
leading to antiselfdual fields on ${\cal M}$.
The canonical quantization of the
action (\ref{prop8}) leads to wavefunctions of the form 
$\exp [i {\cal S}(G) ]$ \cite{nair-schiff}, similar to the usual
connection between the Chern-Simons and WZW theories \cite{witten}.
The action ${\cal S}(G)$ so obtained is on the boundary of 
${\cal M}\times [t_i,t_f]$, i.e.,
on the two equal-time surfaces at $t_i$ and $t_f$ which are copies of ${\cal M}$,
for a compact ${\cal M}$ with no boundary.

The difference with our present situation of
Hall effect is the following.
If we consider not all of ${\cal M}$ but a certain region ${\cal D}$ in
${\cal M}$, there is an additional boundary $\del {\cal D} \times [t_i,t_f]$.
The analogue of the action (\ref{prop7})
on this boundary is what we have in the present case.

\section{Discussion}

We have derived the effective action, in the limit of large number of fermions,
for quantum Hall droplets
in ${\bf CP}^k = SU(k+1)/U(k)$ in the presence of abelian ($U(1)$)
and nonabelian ($SU(k)$)
background magnetic fields. 
This result, which is the main result of this paper, is summarized in
(\ref{63}) and (\ref{63b}).
The action describes a higher dimensional
gauged WZW model.
With the $SU(k)$ nonabelian background field, the states of the lowest Landau level
belong to an $SU(k+1)$ representation $J$, with a set of lowest weight states
forming a representation ${\tilde J}$ of $SU(k)$. The effective action
is for a unitary matrix field $G$ which is an element of
$U(dim{\tilde J})$, which is
coupled to the background $SU(k)$ field
in a left-right symmetric or vectorial way.
There is an $SU(k)$ gauge invariance to take account of this 
coupling to the
$SU(k)$ gauge field. Thus, effectively, the field space is
the coset $U(dim{\tilde J}) /SU(k)$.

The effective action we have derived is very similar to the gauged version of
the K\"ahler WZW model used in the context of higher dimensional
conformal field theories \cite{nair-schiff}. The WZ-term of this action is
defined on the $(2k+1)$-dimensional manifold ${\cal D}\times {\bf R}$,
corresponding to the droplet ${\cal D}$ and time. The rest of the 
action is defined on
$\del {\cal D} \times {\bf R}$, corresponding to 
the boundary of the droplet and time.
The radial variable for the droplet plays the role of the extra dimension
in expressing the WZ-term as the integral of a local differential form.
Further, the action involves the spatial derivative along a particular tangential
direction on the boundary of the droplet, which is eventually integrated over.
There is also a certain chirality
property due to the manner in which the space and time derivatives appear.
Finally, as expected, for
an abelian background, the action reduces to the chiral $U(1)$ bosonic action
given in \cite{KN2}.

The effective action (\ref{63b}) has a very interesting current 
algebra structure
which will have 
implications on the nature of the spectrum of edge excitations. 
This facet of the problem and related issues will be analyzed
in more detail elsewhere.

The WZW theory is well known to lead to conformal field theories
in two dimensions; in particular it is a fixed point under
the renormalization flow. Given that we find a
generalized version of the WZW theory,
it is natural to ask how it changes under renormalization.
We have not carried out a full study of this question, but the following
remark may be interesting in this context.
The variational problem for the action (\ref{1}) gives
the full quantum dynamics. Apart from assuming a Hartree-Fock 
type factorization of
the many-body wavefunctions in terms of one-particle wavefunctions, 
which was explained in \cite{KN2}, we have not
made any other approximation. Thus the effective action we derived 
is the exact quantum action in the large $n$ limit as far as states 
in the lowest Landau level are concerned. Contributions from virtual 
transitions to higher Landau levels have not been calculated.

For a Fermi liquid in dimensions higher than one, expansion around 
the Fermi surface has been used to rewrite the low energy excitations in 
terms of a set of one-dimensional models which can be bosonized. What 
we have done here is similar; our result may be viewed as a partial 
bosonization which is adequate for the low energy excitations.

Finally, we note that while quantum Hall effect and the droplet picture 
may serve as a physical picture for
this class of theories, they can be given an independent formulation.
The Landau levels and the effective field
$G$ are sections of appropriate $H$-bundles on a $G/H$ space.
The kinetic term of the effective action is an analogue
of the Kirillov form. Thus the class of theories we are
discussing may be viewed in a purely geometric way.
It is also clear that these theories may have some relevance to the dynamics
of droplets of incompressible fluids in general.

\vskip .1in
\noindent{\bf Acknowledgements}

This work was supported in part by the National Science Foundation
under grant numbers PHY-0140262 and PHY-0244873 and
by PSC-CUNY grants.

\end{document}